\documentclass[showpacs,amsmath,amssymb,twocolumn]{revtex4}
\pdfoutput=1
\usepackage{subfigure}
\usepackage{graphicx}
\usepackage{dcolumn}
\usepackage{bm}
\usepackage{color}
\usepackage{ulem}

\begin{document}
\title{Glueballs amass at RHIC and LHC Colliders! \\
- The early quarkless 1st order phase transition at $T=270$ MeV\\
 - from pure Yang-Mills glue plasma to GlueBall-Hagedorn states 
        \thanks{This article is based on recent talks on the pure gauge scenario for
the early time dynamics of pp, pA and AA collisions at colliders. 
} 
    } 

\author{Horst Stoecker$^{1,2,3}$}
\author{Kai Zhou$^{2,3}$}
\author{Stefan Schramm$^{2,3}$}
\author{Florian Senzel$^{2}$}
\author{Carsten Greiner$^{2}$}
\author{Maxim Beitel$^{2}$}
\author{Kai Gallmeister$^{2}$}
\author{Mark Gorenstein$^{3,6}$}
\author{Igor Mishustin$^{3}$}
\author{David Vasak$^{3}$}
\author{Jan Steinheimer$^{3}$}
\author{Juergen Struckmeier$^{1,3}$}
\author{Volodymyr Vovchenko$^{1,3}$}
\author{Leonid Satarov$^{3}$}
\author{Zhe Xu$^{4}$}
\author{Pengfei Zhuang$^{4}$}
\author{Laszlo P. Csernai$^{5}$}
\author{Bikash Sinha$^{7}$}
\author{Sibaji Raha$^{8} $}
\author{Tam{\'a}s S{\'a}ndor Bir{\'o}$^{9}$}
\author{Marco Panero$^{10}$}

\affiliation{$^1$GSI Helmholtzzentrum f\"ur Schwerionenforschung GmbH, Planckstra{\ss}e~1, D-64291 Darmstadt, Germany \\ 
$^2$Institut f\"ur Theoretische Physik, Goethe Universit\"at Frankfurt, Max-von-Laue Str. 1, D-60438 Frankfurt Main, Germany \\
$^3$Frankfurt Institute for Advanced Studies, Ruth-Moufang-Str. 1, D-60438 Frankfurt am Main, Germany\\
$^4$Department of Physics, Tsinghua University, Beijing, China\\
$^5$Department of Physics and Technology, University of Bergen, Bergen, Norway\\
$^6$ Bogolyubov Institute for Theoretical Physics 14-b, Metrolohichna str. Kiev, 03680, Ukraine
$^7$VECC, Kolkata, India\\
$^8$Bose Institute, Kolkata, India\\
$^9$Institute for Particle and Nuclear Physics, Wigner RCP, Budapest, Hungary \\
$^{10}$University of Turin and INFN, Turin, Italy
}

\begin{abstract}
The early stage of high multiplicity pp, pA and AA collider is  represented by a nearly quarkless, hot, deconfined pure gluon plasma.
According to pure Yang - Mills Lattice Gauge Theory, this hot pure  glue matter undergoes, at a high temperature, $T_c = 270$ MeV, a first order phase transition into a confined Hagedorn-GlueBall fluid‎. 
These new scenario should be characterized by a suppression of high $p_T$ photons and dileptons, baryon suppression and enhanced strange meson production.
We propose to observe this newly predicted class of events at LHC and ‎RHIC.
\end{abstract}

\date{\today}

\maketitle

The proper understanding of the initial and the early stage of ultra-relativistic pp-, pA- and heavy ion AA- collisions
is a topic of great importance for our understanding of hot and dense QCD matter formed in the laboratory and its phase structure.

At present, the community favors a paradigm of an extremely rapid ($t_{eq}$ less than 0.3 fm/c) thermalization and chemical saturation
of soft gluons and light quarks, their masses and momenta emerging from the decay of coherent massive color flux tubes of strings, which
are formed in the primary hadron-hadron collisions.

However, for a long time also another scenario has been discussed, namely the hot glue scenario, where the initial stage is dominated by
gluons \cite{Shuryak:1992wc,1994PhRvL..73.1895A,1994PhRvD..49.2233M,2001PhRvL..86.1717K}.

We ask the question whether due to initial state color coherence fluctuations two quite distinct classes of
events may exist in collider experiments, or in ultra high energy cosmic ray events, UHECR events. They could be experimentally distinguished in a
high statistics analysis of the collider data at RHIC, LHC, and the FCC, from UHECRs, or from high intensity fixed
target experiments at FAIR \cite{bib3i,bib3j,bib3k,CWBS14,bib5a,bib1d,bib5b,bib3a,bib3l,bib1a,bib1b,bib3d,bib4a,
bib5c,bib3n,bib2c,bib3b,bib3c,bib1c,bib1f,bib1g,bib1i,bib1j,bib2d,bib5g,bib1e,bib1h}, NICA \cite{bib3e} and J-Parc.

Do soft particles at midrapidity in pp-, pA-, and AA- collider experiments develop from an initially quark-free color glass condensate, CGC, through a pre- equilibrium
Glasma-stage into a rapidly chemically saturated, thermalized quarkless pure gluon plasma \cite{XuGreiner(2005)} (see Fig. 1(a)) \cite{Eletal.(2008)}?

\begin{figure*}[t] 
    \begin{center}  
\subfigure[Fugacity of gluons and quarks, respectively.]{ 
\includegraphics[width=0.48\linewidth]{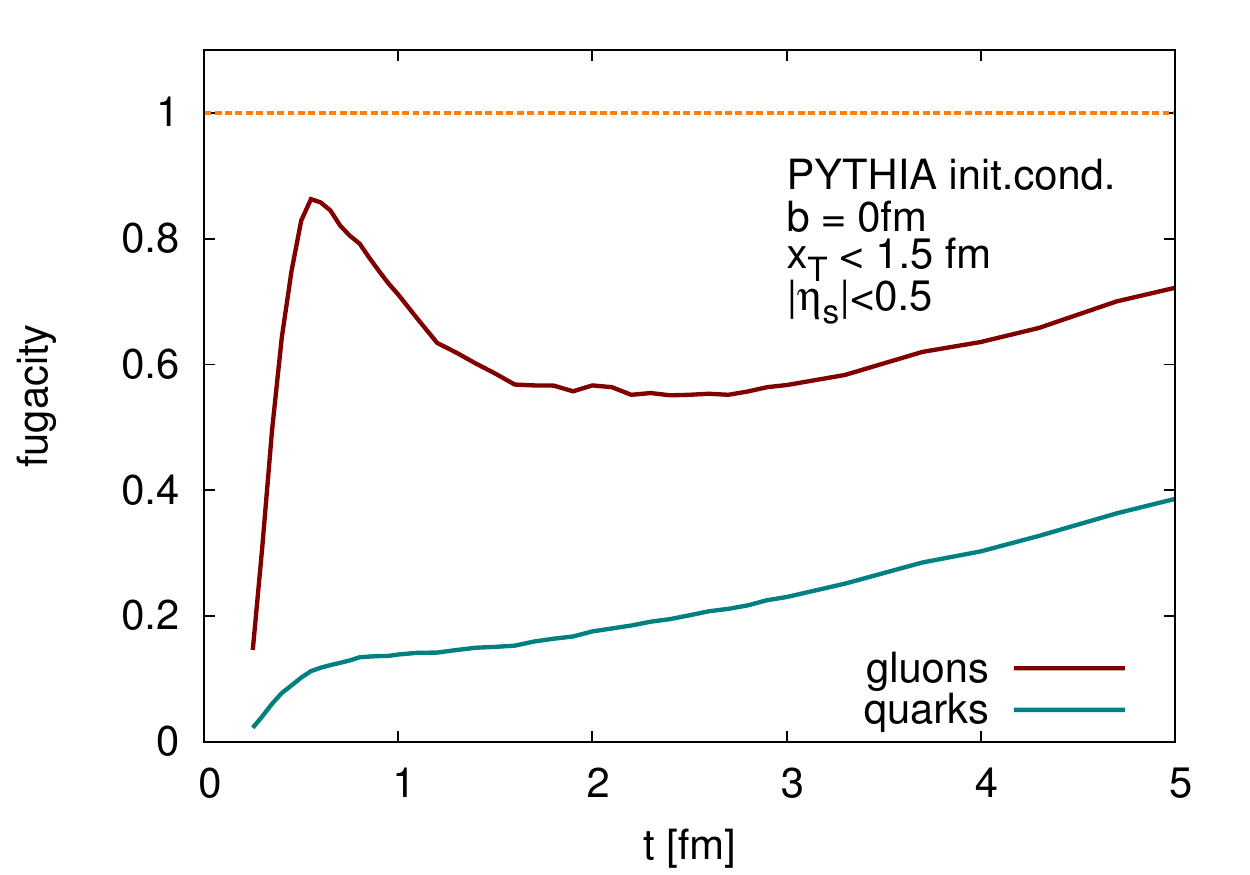} 
}   
\subfigure[number density of gluons and gluon condensate in overpopulated case.]{ 
\includegraphics[width=0.45\linewidth]{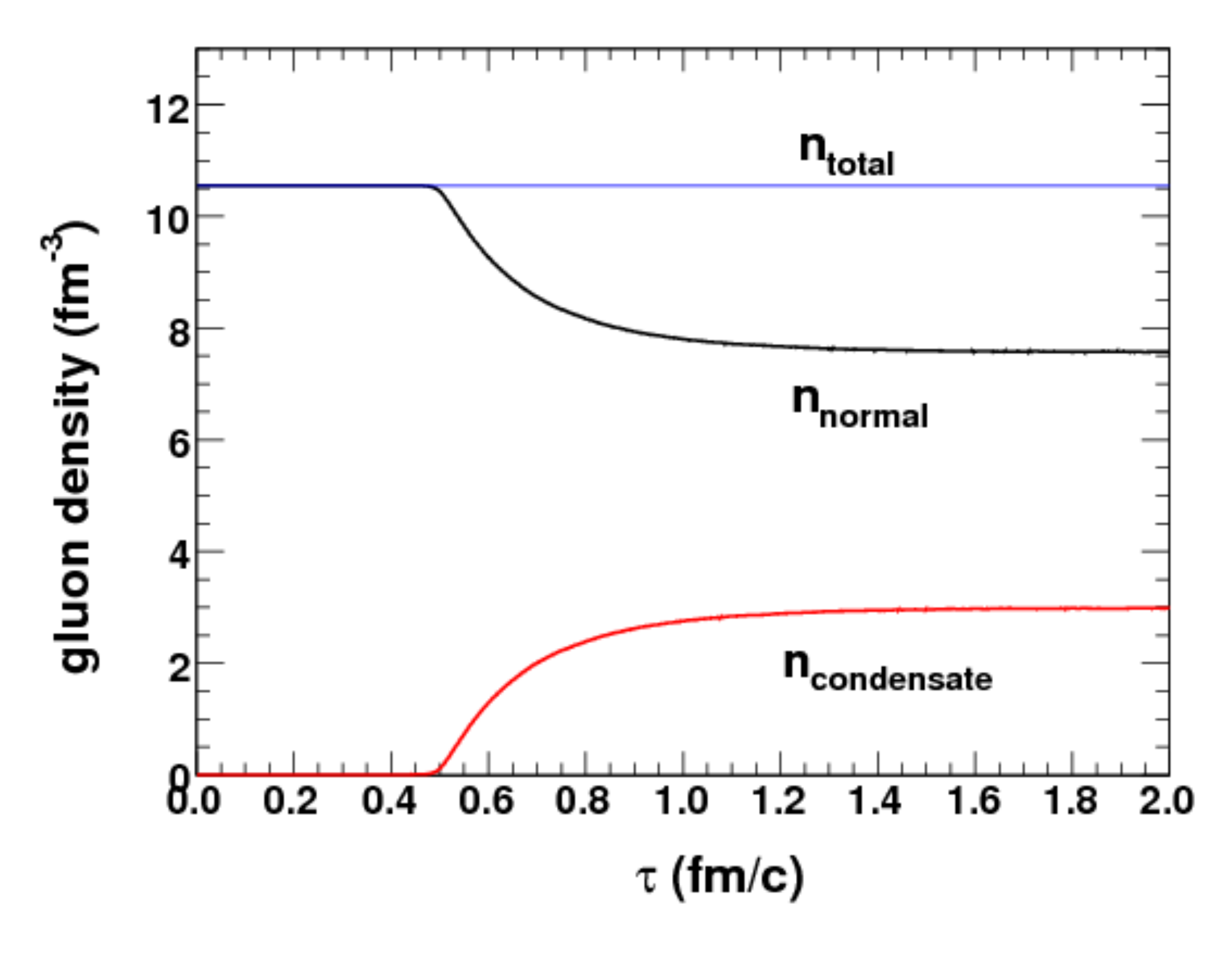} 
}   
\caption{(a) Time evolution for fugacity of gluons and quarks calculated by BAMPS.
Within BAMPS, partons scatter via $2\leftrightarrow2$ interactions in leading-order
pQCD and via inelastic $2\leftrightarrow3$ processes calculated by the improved
Gunion-Bertsch (GB) approximation while considering a running coupling for each
scattering evaluated at the microscopic level. The depicted time evolutions are
calculated for a cylindrical tube of $x_{T} < 1.5$ fm and space-time rapidity
$\eta_{s} < 0.5$ in Au+Au collisions with $\sqrt{s} = 200$ AGeV and an impact
parameter of b=0 fm. (b) Time evolution for number density of gluons and gluon condensate start
with an overpopulated initial condition $f_0=0.4\theta(p-Q_s)$ within BAMPS BOX calculation.} 
    \end{center} 
\label{fig1} 
\end{figure*}

The CGC model predicts that the early Glasma is strongly overpopulated - that means that a
'simple' thermally equilibrated Bose-Einstein distribution can NOT exist, as it can not accommodate the overabundant gluons.

Hence, dynamically a temporary gluon condensate~\cite{jinfeng, zhexu} may be formed in order to accomodate those excess gluons, at least transiently,
see for example Fig. 1(b).

The surprising finding is that only very few soft quarks are present in this early stage according to modern transport
calculations \cite{BMW1992, Roy:1997hj, Elliott:1999uz, 2013NuPhA.904.829cB, 2014NuPhA.930..139B, Uphoffetal.(2015)} (see, however, also \cite{Ruggieri:2015tsa, Scardina:2013nua},
where opposite conclusion of quark equilibration is drawn, mostly due to that they put massive gluons there which can easier produce the lighter quarks, while considering
Debye screening and other non-perturbative chromomagnetic effects). 

As a consequence, only few electric charges are produced, and, thus, there will be hardly any electromagnetic signal from the few soft quarks in the early phase beyond
the photons from hard scattering and direct Drell-Yan dileptons\footnote{Peripheral reactions provide additional possibility to test the initial field 
dynamics.
The intially longitudinal field is of different strength at different transverse points, due
to different colour charges, and the total momentum is also
different \cite{Csernaietal.(1995), Heroldetal.(2013)}. This leads to large angular
momentum \cite{CMSS11}, shear flow and vorticity \cite{CSA12}, with observable consequences.
The CGC field even itself \cite{ChFr13} leads to initial angular momentum 
and vorticity in peripheral collisions. The vorticity is substantial at FAIR 
and NICA energies \cite{CWBS14}, as shown in 3+1d relativistic Particle in Cell
(PICR) model calculations. This vorticity due to equipartition
between local orbital angular momentum and spin, leads to detectable 
$\Lambda$ polarization \cite{XGC15}.
}. 

Yaffe and Svetitski had predicted a sharp first order phase transition in pure SU(3) Yang-Mills gauge
theory \cite{SvetitskyYaffe(1982)}, and modern high accuracy pure LGT results confirm a clear first
order character of this pure gauge phase transition into a confined pure GlueBall
phase \cite{2012JHEP...07..056B,Francis,Saitoetal.(2011),Karsch(2002),Celik:1983wz,Celik:1983nc}.
The critical temperature of this first order transition in pure Yang-Mills SU(3)$_c$ LGT, or quenched SU(3)$_c$, has recently been measured to $T_c = 270$ MeV.
The glueball EOS will affect the expansion rate of the fireball quite significantly.

The presence of a large number of color-confined glueballs in strongly interacting matter must have drastic consequences for the understanding
of the dynamics in high multiplicity pp, pA and AA collisions, as proposed in the pure gauge scenario presented in this paper.

Let us try to recover and understand qualitatively the time evolution of the (initially pure) hot glue system within the so-called ``Columbia plot'' \cite{Brownetal.(1990)}, Fig. 2.

Time dependent effective fugacities, i.e. suppressed light quark numbers, can be toy-modelled by suppression of quark densities
through the use of heavy quark masses, see the nearly diagonal arrow drawn to guide the eye (see Fig. 2):

The first phase of the relativistic collision rapidly creates a thermalized gluon fluid, with initially
'no' (and even up to $t\sim3$ fm/c only very few) soft lighter on-shell quarks and antiquarks present \cite{1994PhRvL..73.1895A}. 
Also the virtual quark- antiquark loops are suppressed by $1/N_c$ \cite{Panero:2009tv, Lucini:2012gg}.
This situation is represented by the upper right hand side corner of the Columbia plot (Fig. 2).

\begin{figure}[t] 
    \begin{center} 
\includegraphics[width=0.8\linewidth]{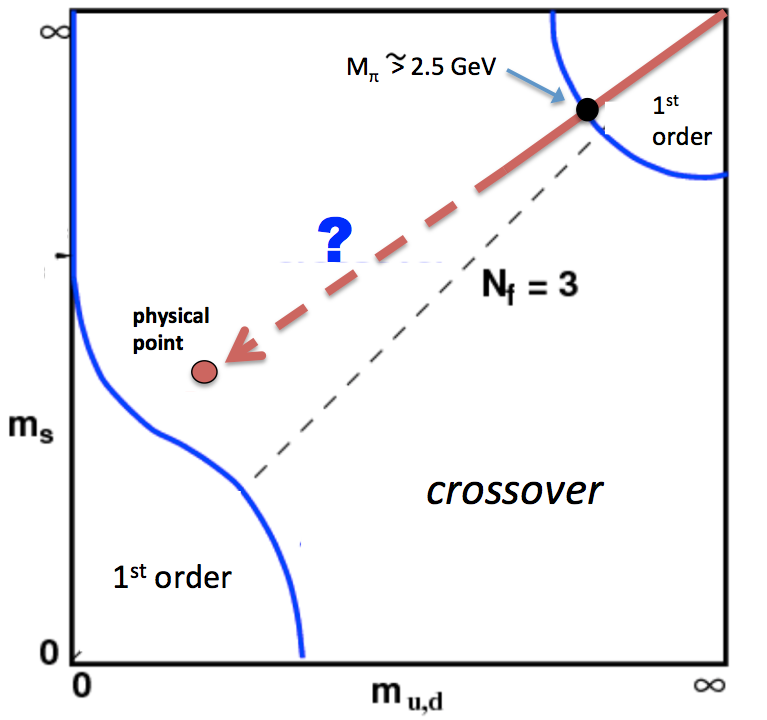} 
    \vspace{0mm} 
    \caption{"Columbia plot" exhibits the dependence of the QCD phase structure, in full equilibrium QCD, on the masses of the different (u,d and s) lighter quarks.
For both, very small and very high, but nearly equal lighter quark masses, QCD exhibits strong first-order phase  transitions, FOPT, between
the confined phase and the deconfined phase, i.e. either between the hadronic and quark-gluon plasma QGP phase, lower lhs corner of
the Columbia plot, and, respectively, between the pure glue plasma phase and the GlueBall phase, upper rhs corner of the plot. 
Early stages of pp, pA and heavy-ion AA collisions are not fully equilibrated, so they can not be represented quantitatively in this plot.
However, the dynamical path of the system may be qualitatively visualized by projecting the time dependent, fugacity weighted composition
and phase structure of the system on the Columbia plot, as it would be resulting from a suppression of the real and virtual lighter quark - antiquark
pair numbers in Lattice Theory. In particular for the early times, $t<3$fm/c, a large effect should be experimentally observable:
Effectively, the change of the QCD phase structure with the different, however small, quark fugacities could easily be numerically estimated by
choosing different, but small, flavor numbers in the lattice simulations. The various 'small $N_f$ value' equations of state, EoS, could then in
turn be used in hydro simulations in order to get better approximations for the time evolution of the system. 
The EoS for the small lighter quark fugacities as calculated on the lattice can be interpolated  between the different
SMALL quark fugacities, as simulated by equilibrium LGT with small $N_f$ values (say $N_f^{effective}= $0.1, 0.3, 0.5, 1.2, 1.8...)} 
    \end{center} 
\label{fig2} 
\end{figure}  

Then, as the system expands, the quarkless, deconfined Yang-Mills matter, hence the pure glue plasma, hits the first order phase transition, at the critical
temperature of 270 MeV, see Fig. 3. Here the deconfined pure glue matter transforms into the confined state of the pure YM theory, i.e. into a GlueBall fluid.
This surprising prediction holds as long as there are not sufficiently many (virtual and real) quarks formed, as in this situation a
(possibly even supersaturated) deconfined hot pure gluon plasma has only one confined exit channel, namely forming a glueball fluid.

Entropy conservation (adiabatic cooling and expansion) forces the volume of the system to expand at the constant critical
temperature $T_c$ of the phase transition, and only after all glue plasma has completely transformed
into the GlueBall fluid, at $T_c=270$ MeV, will it be possible to cool the system further: possibly
the heavy glueballs in this dilute (!) glueball fluid collide and decay - into lighter GlueBalls,
but this is the end of the pure gauge story. 

This new pure Yang-Mills scenario, with its radically different collision history is sketched in Fig. 3.

\begin{figure}[tp] 
    \begin{center} 
\includegraphics[width=0.9\linewidth]{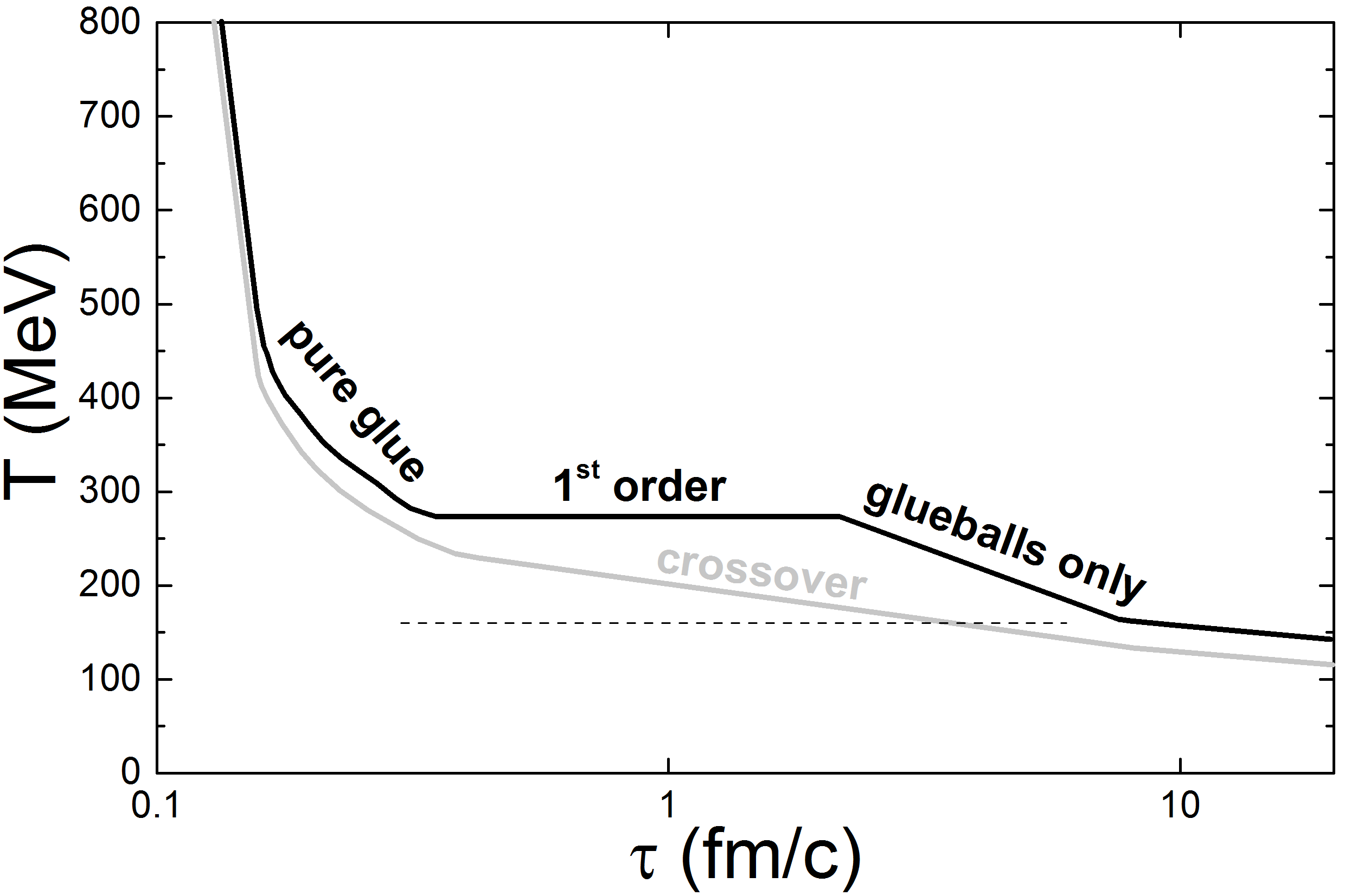} 
    \vspace{0mm} 
    \caption{Sketch of the time evolution of a high-energy collision in the pure glue scenario with Yang-Mills 1sr order phase transition to GlueBalls.} 
    \end{center} 
\label{fig3} 
\end{figure}  

In a more realistic description of pp, pA and AA collision, one shall include the fact that some quarks
will already be produced before and during the first order phase transition, FOPT \cite{vovchenko_tobe}.

This could be modeled by moving from the upper right hand side of the Columbia plot along the diagonal
to effectively less heavy quark masses in the plot, as indicated in Fig. 2, until the boundary
line of the upper r.h.s. quadrant, which indicates that the phase transition changes its order from first to second
order, is reached. This will happen at a critical temperature, $T_{2nd order}\simeq 204$ MeV, about $10\%$ below the pure gauge $T_c=270$ MeV \cite{Borsanyi:2012ve}. 

But even beyond this transition line, in the crossing region of the Columbia plot, the lower temperature confined phase
is heavily populated by GlueBalls and GlueBall-Hagedorn resonances \cite{Meyer:2004jc, Meyer:2008tr, Borsanyi:2012ve}:

Recently, an increasing number of lattice calculations have studied GlueBalls both in pure SU(3) gauge
theory \cite{2014arXiv1411.3503M,1999PhRvD..60c4509M,Meyer:2004jc,2006PhRvD..73a4516C,Caselle:2011fy, Caselleetal.(2015)} as well as in
full 2+1 flavor LQCD, reasonably close to the physical point \cite{2010PhRvD..82c4501R,2012JHEP...10..170G,2012JHEP...07..056B,Borsanyi_etal.(2014)}.
Although the dynamical quarks in full 2+1 flavor LQCD introduce some hadronic mixing (see, for example, \cite{Ochs:2013gi})
and therefore change the masses of the GlueBalls as measured on the lattice, it can be stated that these peculiar,
(mostly) fermion-free confined heavy hadrons are firmly predicted by the theory of strong interactions (see also \cite{Hessetal.(1999),Mishustin:2006wm,Philipsen:2012nu}).

On the other hand, the study of unquenched QCD thermodynamics and the associated phase transition with small
quark masses has made substantial progress \cite{Borsanyi_etal.(2014),Bazavovetal.(2014),Bhattacharyaetal.(2014),Brandtetal.(2012),Brandt(2013)}.

Combining both research thrusts, the role of the GlueBalls for the equation of state of strongly interacting matter has gained
quite some interest lately \cite{Meyer(2009),Mengetal.(2009),GavaiGupta(2000),Caselle:2011fy,Caselleetal.(2015)}.
Thermal lattice QCD calculations have found substantial contributions from glueballs to pressure and entropy of the hadron
resonance gas HRG, both below and above the crossing transition from deconfined to confined matter,
and strong evidence for exponentially increasing Hagedorn towers of GlueBalls has been presented \cite{Meyer(2009),Borsanyi_etal.(2014)}-see also\cite{Buisseret:2011fq}.
The hadron resonance gas model HRG needs to incorporate not only the measured states from the PDG, but also states which are new (X, Y, Z, pentaquarks,
four quark states and/or other exotica as predicted by QCD/LQCD, like glueballs). For refs see eg \cite{Meyer:2015eta}.

The GlueBalls and the heavy GlueBall-Hagedorn resonances can decay rapidly in a consecutive cascade, a chain of two body decays,
into light and heavy Hagedorn states \cite{Beitel:2014kza}. The bulk of the hadronic final state will be produced from the
lighter GlueBall Hagedorn states which decay finally into hadronic resonances and light hadrons. Not too surprising is the
fact that both the final hadronic yields and the slopes of the spectra do approach the experimental data in a natural way,
as predicted by Hagedorn decades ago, and as recently shown by Beitel et al., by comparing their covariant Frautschi - Hagedorn
modell directly to the RHIC and LHC data of the STAR and ALICE collaborations.

As the temperature of the pure gauge first order phase transition, FOPT, is much higher, $T_c\sim 270 MeV$, than the crossover
temperature, $T_{cr}$, of full LQCD at the 2+1 flavor 'physical point' of a fully equilibrated, thermalized quark-gluon system,
$T_{cr} = 155 \pm 20$ MeV, this idealized (= infinitely slow transition) 'physical point' of full 2+1 flavor hot QCD may not even
be reached in RHIC and LHC at high multiplicity pp, pA, and also not in the initial stage AA collisions (See Fig. 4), as the GlueBall states
may decay directly from a higher temperature phase.

\begin{figure}[t] 
    \begin{center} 
\includegraphics[width=1.0\linewidth]{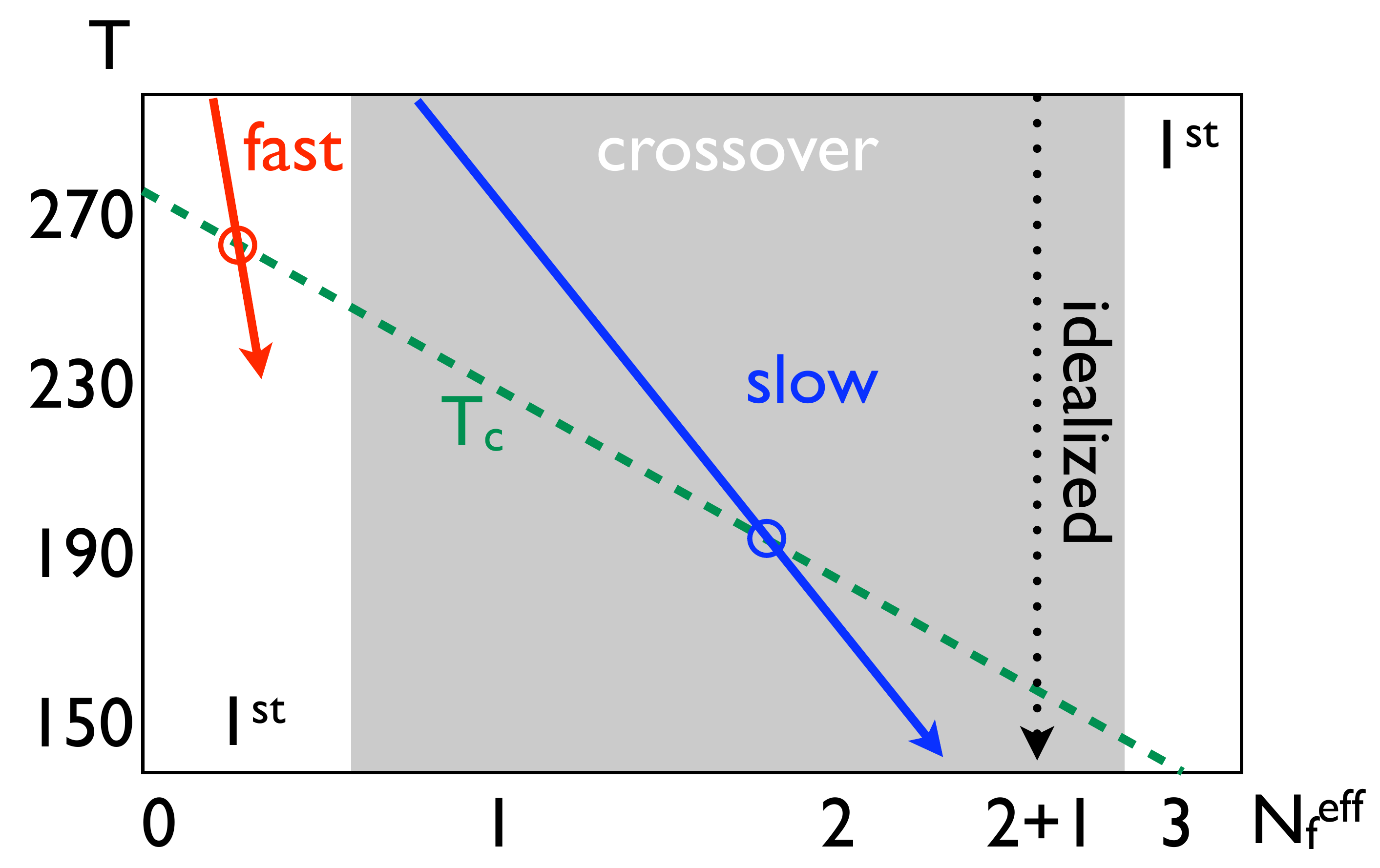} 
    \vspace{0mm} 
    \caption{Transition temperature versus the effective number of flavours.} 
    \end{center} 
\label{fig4} 
\end{figure}

One interesting aspect to investigate would be the dynamics of passing through the first-order gluonic transition.
As has been discussed in \cite{CsernaiKapusta(1992)}, passing a 1st order transition might take a long time. A sudden transition of a
supercooled state, however, could speed up this process significantly \cite{Csernaietal.(1995)}. To our knowledge no quenched
lattice QCD calculation exists that studies the range of temperatures where supercooling and -heating is possible,
i.e., determining the parameters of the spinodal instabilities (see some early attempts in \cite{Callaway:1982eb,Polonyi:1983tm,Polonyi:2005cq,Kogut:1984sa}). Such a calculation would be very relevant in this context.
Here we show pure glue hydro simulations \cite{Steinheimer.future}: Fig. 5 depicts gluon density along z-axis in a supercooled pure glue hydrodynamic
expansion and Fig. 6 shows bubble formation in supercooled rapid phase transition of pure glue model.
\begin{figure}[tp] 
    \begin{center} 
\includegraphics[width=1.0\linewidth]{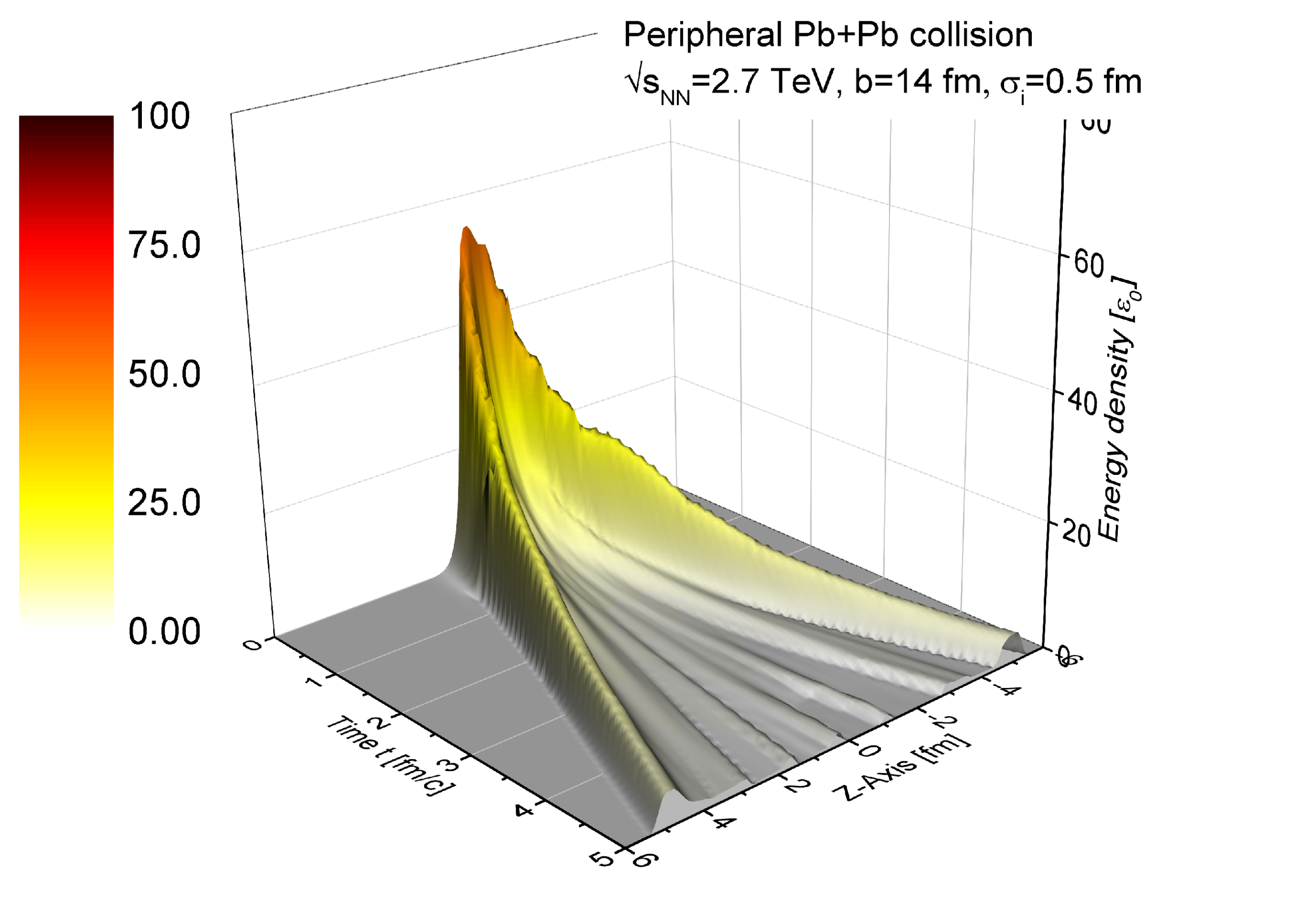} 
    \vspace{0mm} 
    \caption{Gluon density along with z-axis in supercooled pure glue hydrodynamic expansion.} 
    \end{center} 
\label{fig5} 
\end{figure}

\begin{figure}[tp] 
    \begin{center} 
\includegraphics[width=1.0\linewidth]{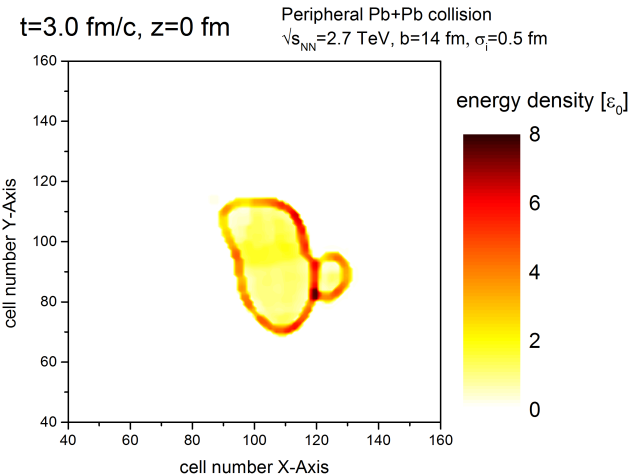} 
    \vspace{0mm} 
    \caption{Bubble formation in supercooled rapid phase transtion of pure glue model.} 
    \end{center} 
\label{fig6} 
\end{figure}

Pisarski and Wilzcek had predicted already in the 1980ies that fully thermalized QCD with massless quarks (in the chiral limit) will
exhibit a soft crossing rather than a sharp first order transition, which has been confirmed by numerous lattice simulations
lately \cite{1984PhRvD..29..338P,Borsanyi_etal.(2014),Bazavovetal.(2014),Bhattacharyaetal.(2014),Brandtetal.(2012),Brandt(2013)},
see also the review by O. Philipsen \cite{Philipsen:2012nu}.

The success of the hydrodynamic model in describing the bulk of hadronic production in Au+Au collisions at RHIC has led to paradigmatic
shift in our view of the QGP: instead of behaving like a gas of weakly interacting quarks and gluons, as naively expected on the basis of
asymptotic freedom in QCD, it’s collective properties rather reflect those of a ”perfect fluid” with almost vanishing
viscosity at the crossing transition \cite{Stoecker:1986ci, Kodama:95652, Song:2007fn, Stoecker:2004qu, Steinheimer:2007iy}. Actually
it is the most perfect fluid created in the laboratory. It is also highly opaque to colored probes, as indicated by the observed large parton
energy loss. Actually these two phenomena are  fundamentally related to each other, and complement each other to bring about a strongly
coupled plasma \cite{Heinz:2015lpa}.

However, as the virtual quark loops are suppressed like $1/N_c$ and soft lighter quarks, which could screen the gluon interaction and
therefore push down the critical temperature, are not yet produced in appreciable numbers in the first stage of the systems time
evolution, $t < 3$fm/c (which in turn also means that, $N_f\ll 1$ at $t<3$ fm/c, i.e. the effective number of flavors is small), the often used assumption of an
immediate fully saturated thermal and chemical equilibration, $t=0$, which would allow the use a full $N_f = 2+1$ QCD EoS in
hydro calculations, is - at least during the early, $t < 3$fm/c, stage of the time evolution of the system, not justified.
Therefore, as alternative paradigm, the pure glue scenario, $N_f\sim0$, as discussed above, is more appropriate to study the
fate of the system for the early time evolution, and should also be applied in hydrodynamic modelling of the collision - with
time dependent fugacities until quark saturation is reached, and a phase structre and an equation of state appropriate for the small time dependent effective
flavor number $N_f(t)$ \cite{vovchenko_tobe}.    

Thus, the pure gauge theory scenario as outlined above, if realized in nature, may prevent us from seeing the lower temperature crossing at
$T_{cd}\sim 155$ MeV, due to the small quark number and
small net quark densities in high multiplicity pp and pA events and in the early phase of peripheral (and possibly even in the early phase of central) AA collisions. 

If such a situation occurs in current laboratory experiments this is truely a fortiori in the extensive air showers initiated by ultrahigh-energy cosmic
rays (UHECR). The initial interaction of the cosmic-ray particle with the atmosphere represents essentially a nucleon-nucleus or nucleus-nucleus
collisions at energies much higher than what can be achieved in man made experiments \cite{2014CRPhy..15..318K}. The projectile energies are
$6\times10^{19}$eV at the GZK cutoff \cite{1966PhRvL..16..748G,1966JETPL...4...78Z}. At these yet higher energies, the relative dominance of
the gluons over the quarks should be even more pronounced.

If the Higgs sector is only a low-energy manifestation of a new strongly interacting gauge sector, these comic-ray energies might even be
sufficient to trigger a similar scenario in this new gauge sector \cite{Dietrich:2012vi}. For technicolor, UHECR energies are what RHIC energies are for QCD.

The pure glue scenario is expected to occur also, however for a fleeting moment only, in the early universe: At the QCD phase transition, $10^{-5}$ seconds
into the expansion of the universe, the dynamics allows the quarks to catch up with the gluons. 

Likewise, at the electroweak phase transition one may search for the pure gauge transition, if the expansion rate of the universe is not too slow to allow
for a gauge-boson dominated phase transition as a strongly interacting mechanism for electroweak symmetry breaking.

{\bf Conclusions:} The early stage of high multiplicity pp, pA and AA collider events can represent a new class of events: a nearly quarkless, hot, deconfined pure gluon plasma.
According to pure Yang - Mills Lattice Gauge Theory, this hot pure  glue matter undergoes, at a high temperature, $T_c = 270$ MeV, a first order phase transition into a confined Hagedorn-GlueBall fluid‎. 
These events should be characterized by a suppression of high $p_T$ photons and dileptons, baryon suppression and enhanced strange meson production.
We propose to observe this newly predicted class of events at LHC and ‎RHIC.

{\bf Acknowledgements:} We would like to acknowledge valueable contributions by
Larry McLerran, Owe Philipsen, Christian Sturm, Paolo Guibellino, John Harris, Helmut Oeschler, Anton Andronic, Silvia Mascciotti,
Zoltan Fodor, Szabolcs Borsanyi, Frithjof Karsch, Karl-Heinz Kampert, Dennis D. Dietrich, Krzysztof Redlich,
Bengt Friman, Peter Braun-Munziger, Ajit M. Srivastava, Francois Arleo,  Wolfgang Cassing, Elena Bratkovskaya and Joerg Aichelin.

\end{document}